\documentclass[12pt]{article}
\usepackage{setspace}
\usepackage{caption}
\usepackage{amsfonts}
\usepackage{graphicx}
\usepackage{color}
\usepackage{amsmath}
\usepackage{float}
\usepackage{comment}
\newcommand {\be}{\begin{equation}}
 \newcommand {\ee}{\end{equation}}
 \newcommand {\bea}{\begin{array}}
 
 \newcommand {\eea}{\end{array}}

\textwidth=6.5in \hoffset=-.75in \voffset=-1in \numberwithin{equation}{section}
\numberwithin{figure}{section}

\begin{document}

\begin{titlepage}
\vspace{1cm} 
\begin{center}
{\Large \bf {Destroying Kerr-Sen black holes}}\\
\end{center}
\vspace{2cm}
\begin{center}
\renewcommand{\thefootnote}{\fnsymbol{footnote}}
Haryanto M. Siahaan{\footnote{haryanto.siahaan@unpar.ac.id}}
\\
Physics Department, Parahyangan Catholic University,\\
Jalan Ciumbuleuit 94, Bandung 40141, Indonesia

\renewcommand{\thefootnote}{\arabic{footnote}}
\end{center}

\begin{abstract}
By neglecting the self-force, self-energy, and radiative effects, it has been shown that an extremal or near-extremal Kerr-Newman black hole can turn into a naked singularity when it captures charged and spinning massive particles. A straightforward question then arises: do charged and rotating black holes in string theory possess the same property? In this paper we apply the Wald's gedanken experiment, in his study on the possibility of destroying extremal Kerr-Newman black holes, to the case of (near-)extremal Kerr-Sen black holes. We find that feeding a test particle into a (near-)extremal Kerr-Sen black hole could lead to a violation of the extremal bound for the black hole. 

\end{abstract}
\end{titlepage}\onecolumn 
\bigskip 

 \section{Introduction}
\label{sec:intro}

According to the cosmic censorship hypothesis, all physical singularities due to gravitational collapse are hidden behind an event horizon \cite{Penrose,Joshi:2008zz}. This hypothesis, which implies that no naked singularity occurs in our Universe, is formulated in weak and strong versions \cite{Joshi:2008zz}. The weak cosmic censorship conjecture (WCCC), which has a relevance to the work presented in this paper\footnote{The weak censorship conjecture deals with the asymptotically flat spacetime \cite{Joshi:2008zz}, which is a feature in Kerr-Sen geometry.}, effectively postulates that the singularities due to gravitational collapse cannot influence points near the future null infinity ${\cal I}^+$. For example, in the case of collapsing stars the hypothesis says that the singularity resulting from this process must be hidden behind an event horizon. However, due to the lack of solid evidence that a black hole candidate is really a black hole, the possibility of a naked singularity's existence is worth considering. Related to this consideration, it is interesting to note that one can observationally differentiate naked singularities from black holes through the characteristics of their gravitational lensings \cite{Virbhadra}. 

In the Einstein-Maxwell theory, several investigations on the WCCC violation have been carried out in the literature. For example, in his groundbreaking work \cite{Wald} Wald showed that it is impossible to turn an extremal Kerr-Newman black hole into a naked singularity by letting the black hole capture a test particle having large angular momentum and electric charge compared to its energy. Later on, the problem of WCC violation was revisited by many authors; for example in \cite{Hubeny:1998ga} Hubeny showed that overcharging a near-extremal Reissner-Nordstrom (RN) black hole is possible by injecting charged test particle into the black hole, and in \cite{Jacobson:2009kt} Jacobson and Sotiriou showed that a near-extremal Kerr black hole can be overspun by a test particle with angular momentum. 

Quite recently Saa et al. in \cite{SaaPRD} showed that, by neglecting the backreaction effect, destroying a near-extremal Kerr-Newman black hole is possible by a test particle with electric charge and angular momentum. In their analysis, the particle's energy is kept linear in the equation related to the extremality, and the overextremization of Kerr-Newman black holes cannot be performed once the black holes are in the extremal condition. Later on, Gao et al. showed in \cite{Gao:2012ca} that, by neglecting the radiative and self-force effects, destroying an extremal Kerr-Newman black hole with a test particle is possible if the linear approximation of the particle's energy is not taken into account. However, due to the narrow range of the particle's energy, which leads to the violation of the black hole's extremality bound, taking the radiative and self-force effects \cite{CardosoPRL,CardosoPRD,Zimmerman:2012zu} into account could be a cure to the problem of producing a naked singularity from a black hole.

Several studies in literature about the possibility of cosmic censorship violation in charged and/or rotating black holes are also worth mentioning. In \cite{Semiz}, the authors studied the possibility of violating WCCC in the case of a black hole interacts with fields instead of test particles. The possibility of producing a naked singularity in a Kerr-Newman background by letting a neutral spinning body fall into an extremal RN black hole was discussed in \cite{Felice}. Keeping up to the linear order in the test particle parameters, extremal black holes can at most remain extremal in a variety of scenarios \cite{LinearRemain}. Including the cosmological constant in studies of WCCC violation of black holes was considered in \cite{WCC-cosmo}. Very recently, a nonperturbative test of cosmic censorship with a stream of
charged null dust in the theory as discussed in the present paper was performed in \cite{Pani}, where the authors showed that some energy conditions prevent the formation of a naked singularity in the future. 

In the low energy limit of string field theory, there is a known rotating charged black hole solution, namely the Kerr-Sen black hole \cite{sen}. It has physical properties which are quite similar to those in Einstein-Maxwell theory, but it can still be distinguished in several aspects. For example, the authors of \cite{Hioki:2008zw} studied the capturing and scattering of photons in the background of Kerr-Sen and Kerr-Newman black holes, where some characteristic differences of the capture region due to their spacetime structures are presented. Also, the authors of \cite{Koga:1995bs} discussed the evaporation process of these black holes and found their emission rates to be distinguishable, and from the work by Horne and Horowitz \cite{Horne:1992zy} we learn that the gyromagnetic ratios of Kerr-Newman and Kerr-Sen black holes are different. In addition to these examples, the hidden conformal symmetries of Kerr-Sen black holes were studied in \cite{KSKN}, where the authors found that Kerr-Sen black holes do not have the $Q$-picture hidden conformal symmetry that Kerr-Newman black holes possess \cite{Chen:2010ywa}. Some similar properties, for example, are the instability of bound state charged massive scalar fields in these black hole backgrounds \cite{Nambu:2014,Siahaan:ijmpd2015} and also the CFT$_2$ holographic dual for the scattering process in the background of these black holes \cite{Ghezelbash:2014aqa}. 

Motivated by several resemblances between the physical properties of Kerr-Newman and Kerr-Sen black holes, we would like to study the possibility to turn a (near-)extremal Kerr-Sen black hole into a naked singularity by adopting the method of testing WCCC for (near-)extremal Kerr-Newman black holes as in \cite{SaaPRD,Gao:2012ca}. We show that, by neglecting the radiative and self-force effects, the WCCC is violated for Kerr-Sen black holes, in both extremal and near-extremal cases. By using the outcomes in our study of WCCC violation for Kerr-Sen black holes, we discuss naked singularity production from extremal Kerr and Gibbons-Maeda-Garfinkle-Horowitz-Strominger (GMGHS) black holes \cite{GMGHS}, i.e., the limits $Q\to 0$ and $a\to 0$ of Kerr-Sen black holes respectively. We also provide a numerical plot showing that a test particle, which potentially could overspin and/or overcharge the black hole, could really fall all the way from $r\to \infty$ into the black hole. 

The organization of this paper is as follows. In Sec. \ref{s.1} we review some properties of Kerr-Sen black holes. Then in Sec. \ref{s.2} we obtain the constants of motion $E$ and $L$, i.e. the energy and angular momentum of the test particle, in Kerr-Sen background respectively. Subsequently, in Sec. \ref{s.3} we show how a (near-)extremal Kerr-Sen black hole can be destroyed by an infalling test particle. In Sec. \ref{s.4} we use the result obtained for Kerr-Sen to study the production of naked singularities from Kerr and GMGHS black holes. Finally in Sec. \ref{s.discussion}, we give a discussion and our conclusion. In this paper we use the unit system where $c = G = \hbar = 1$. 

\section{Kerr-Sen black holes}\label{s.1}

In Ref. \cite{sen}, Sen obtained a four-dimensional solution that describes a rotating and electrically charged massive body in the low energy heterotic string field theory. The corresponding effective action in this theory reads \cite{sen}
\be \label{action-low-het}
S = \int {d^4 x\sqrt {\left| {\tilde g} \right|} } e^{-\tilde{\Phi}}\left( {R - \frac{1}{8}F^2  + {\tilde g}^{\mu \nu } \partial _\mu  \tilde \Phi \partial _\nu  \tilde \Phi   - \frac{1}{{12}}H^2 } \right)\,,
\ee
where ${\tilde g}$ is the determinant of metric tensor ${\tilde g}_{\mu\nu}$; $F^2$ is the square of field-strength tensor $F_{\mu \nu }  = \partial _\mu  A_\nu   - \partial _\nu  A_\mu  $; $\tilde{\Phi}$ is the dilaton; $H^2$ is the square of a third rank tensor field:
\be
H_{\kappa \mu \nu }  = \partial _\kappa  B_{\mu \nu }  + \partial _\nu  B_{\kappa \mu }  + \partial _\mu  B_{\nu \kappa }  - \frac{1}{4}\left( {A_\kappa  F_{\mu \nu }  + A_\nu  F_{\kappa \mu }  + A_\mu  F_{\nu \kappa } } \right)\,,
\ee 
and $B_{\mu\nu}$ is a second rank antisymmetric tensor field. It is obvious that in the case all nongravitational fields in the action (\ref{action-low-het}) vanish, we get the Einstein-Hilbert action. Since Kerr metric solves the vacuum Einstein equations, it is also a solution in the theory described by (\ref{action-low-het}) when all the nongravitational fields are absent. In fact, the Kerr-Sen solution \cite{sen} is obtained by applying a set of transformations which connects the solutions in (\ref{action-low-het}) to the Kerr metric.

Now let us review some aspects of the Kerr-Sen solution. In the Boyer-Lindquist coordinates $(t,r,\theta,\phi)$, the nonvanishing components of Kerr-Sen tensor metric in the Einstein frame\footnote{The spacetime metric in the string frame ${\tilde g}_{\mu\nu}$ and Einstein one $g_{\mu\nu}$ are related by $g_{\mu \nu }  = e^{ - \tilde \Phi } {\tilde g}_{\mu \nu } $.} are
\[{g_{tt}  =  - \frac{{\Delta  - a^2 \sin ^2 \theta }}{{\rho ^2 }}
~~,~~g_{t\phi }  = -\frac{{2Mra\sin ^2 \theta }}{{\rho ^2 }}}
\]
\be \label{KerrSen-metric}
{g_{rr}  = \frac{{g_{\theta \theta } }}{\Delta } = \frac{{\rho ^2 }}{\Delta }
~~,~~
g_{\phi \phi }  = \sin ^2 \theta \left( {\Delta  + \frac{{2Mr\left( {r\left( {r + 2b} \right) + a^2 } \right)}}{{\rho ^2 }}} \right)}\,,
\ee 
where $\rho^2=r(r+2b)+a^2\cos^2\theta$, $\Delta=r(r+2b)-2Mr+a^2$, and $b=Q^2/2M$. The rotational parameter $a$ is defined as a ratio between the black hole's angular momentum $J$ to its mass $M$. The black hole's electric charge is denoted by $Q$, and the spacetime described by the Kerr-Sen metric is not vacuum, analogous to the Kerr-Newman case in the Einstein-Maxwell theory. The solutions for nongravitational fundamental fields in the theory described by the action (\ref{action-low-het}) are \cite{sen}
\begin{eqnarray}
{\tilde\Phi}&=&-\frac{1}{2}\ln \frac{\rho ^2}{{r}^2+a^2\cos^2\theta},\label{dil}\\
A_{{t}}&=&\frac{-{r}Q}{\rho ^2},\label{A-t}\\
A_{{\phi}}&=&\frac{{r}Qa\sin^2\theta}{\rho ^2},\label{A-phi}\\
B_{{t}{\phi}}&=&\frac{b{r}a\sin^2\theta}{\rho ^2}.\label{antisym-tensor-field}
\end{eqnarray}

The line element (\ref{KerrSen-metric}) contains black hole solution whose outer and inner horizons are located at $r_+=M-b+\sqrt{(M-b)^2-a^2}$ and $r_-=M-b-\sqrt{(M-b)^2-a^2}$ respectively. The corresponding Hawking temperature, angular velocity, and electrostatic potential at the horizon are, respectively, given by
\begin{eqnarray}
T_H&=&\frac{r_+-r_-}{8\pi Mr_+} =\frac{\sqrt{(2M^2-Q^2)^2-4J^2}}{4\pi M(2M^2-Q^2+\sqrt{(2M^2-Q^2)^2-4J^2})},\label{TH} \\
\Omega_H&=& \frac{a}{2Mr_+} =\frac{J}{M(2M^2-Q^2+\sqrt{(2M^2-Q^2)^2-4J^2})},\label{Omeg}\\
\Phi_H&=&\frac{Q}{2M}\label{Electrostatic_H}.
\end{eqnarray}
Setting the parameter $b=0$ yields all nongravitational fields (\ref{dil}) - (\ref{antisym-tensor-field}) vanish, and the line element (\ref{KerrSen-metric}) reduces to the Kerr metric. Furthermore, turning off the rotational parameter $a$ in (\ref{KerrSen-metric}) which is followed by the $r \to r-2b$ shift give us the GMGHS solution\cite{GMGHS}, which describes the spacetime outside of an electrically charged mass in string theory.

\section{Energy and angular momentum of test particles}\label{s.2}
In a general curved background, the motion of a massive charged test particle is dictated by the equation \cite{Misner:1974qy}
\be\label{eq.motion.incurved}
{\ddot x}^\mu + \Gamma _{\alpha \beta }^\mu {\dot x}^\alpha {\dot x}^\beta   = \frac{q}{m}F^{\mu \nu } {\dot x}_\nu\,,
\ee 
where $m$ and $q$ are the mass and electric charge of the particle respectively. In the equation above, the ``dot'' stands for the derivative with respect to some affine parameter $s$, ${\dot ()} = \tfrac{d()}{ds}$. A Lagrangian which yields the equation (\ref{eq.motion.incurved}) can be expressed as \cite{Wald}
\be\label{eq.Lang}
{\cal L} = \frac{1}{2}mg_{\alpha \beta } {\dot x}^\alpha {\dot x}^\beta + qA_\mu  {\dot x}^\mu\,.
\ee 
Accordingly, the energy $E$ and angular momentum $L$ as the constants of particle's motion in a stationary background are given by
\be
E =  - \frac{{\partial {\cal L}}}{\partial {\dot t}} =  - m\left( {g_{tt} \dot t + g_{t\phi } \dot \phi } \right) - qA_t\, ,
\ee 
and
\be 
L = \frac{{\partial {\cal L}}}{\partial {\dot \phi}} = m\left( {g_{t\phi } \dot t + g_{\phi \phi } \dot \phi } \right) + qA_\phi
\ee 
respectively. 

Due to the timelike condition for the charged massive particle under consideration, ${\dot x}_\mu{\dot x}^\mu = -1$, one can obtain a relation between $E$ and $L$ from the last two equations above which takes the form
\be \label{EtoL}
E = \frac{{g_{t\phi } }}{{g_{\phi \phi } }}\left( {qA_\phi   - L} \right) - qA_t  + {\sqrt {{\left( {\frac{{g_{t\phi }^2  - g_{\phi \phi } g_{tt} }}{{g_{\phi\phi }^2 }}} \right)\left( {\left( {L - qA_\phi  } \right)^2  + m^2 g_{\phi \phi } \left( {1 + g_{rr} \dot r ^2 + g_{\theta \theta } \dot \theta ^2} \right)} \right)}}}\,.
\ee
In getting the last equation we have considered the solution that implies ${\dot t} > 0$ only. Furthermore, inserting the corresponding components of $g_{\mu\nu}$ in (\ref{KerrSen-metric}) into (\ref{EtoL}) which is followed by the $r=r_+$ evaluation, we find the minimum energy which allows the test particle to (classically) reach the event horizon as
\be\label{Emin-notEx}
E_{\rm min} = \frac{{aL + qQr_ +  }}{{2Mr_ +  }} \,.
\ee 
Consequently, we now have a lower bound for the particle's energy which could destroy the black hole's horizon, $E \ge E_{\rm min}$.

\section{Kerr-Sen black holes and naked singularity}\label{s.3}

Kerr-Sen black holes become extremal when $M = |a| + b$, where the inner and outer horizons coincide with each other. By capturing a massive and charged test particle, the reading of the extremal condition becomes
\be\label{extremaleqtn}
2\left( {M + E} \right)^2  = 2\, |J+L| + \left( {Q + q} \right)^2 \,,
\ee
where the test particle's energy, angular momentum, and electric charge are denoted by $E$, $L$, and $q$, respectively. For the sake of simplicity in writing the formulas, from now on we chose not to write the absolute signs for $L$, $J$, and $a$ explicitly, and we assume that $J$ and $L$ have the same sign, and that $L$, $J$, and $a$ all have positive values. Now we consider the near-extremal condition, indicated by $M - a - b = \delta$, where $0 < \delta \ll M$. Accordingly, the near-extremal Kerr-Sen black hole turns to a naked singularity if
\be\label{delEMQ}
\delta M + 2ME + E^2  < Qq + L + \frac{{q^2 }}{2}\,.
\ee 

In \cite{SaaPRD}, the authors considered up to the linear term of $E$ only in the equation analogous to (\ref{delEMQ}) for Kerr-Newman black holes. To get a more precise upper limit for the particle's energy which allows violation of black hole's extremal bound, we keep the nonlinear term of $E$ in (\ref{delEMQ}). In this scheme we obtain 
\be \label{E-max}
E_{\rm max} = M(X-1)
\ee
as the upper bound for test particle's energy where $X = \sqrt{1+\left({L+q(Q+q/2)-\delta M}\right)/M^{2}}$. It is the existence of $\Delta E \equiv E_{\max }  - E_{\min }  > 0$, where $E_{\max }$ and $ E_{\min }$ are given in (\ref{E-max}) and (\ref{Emin-notEx}) respectively, that indicates the possibility of producing a naked singularity when a near-extreme Kerr-Sen black hole captures a massive charged particle. 

For the near-extremal Kerr-Sen black holes, the outer horizon radius can be written as $r_+ =a+\delta+\sqrt{\delta(\delta+2a)}$, which therefore yields
\be \label{eqDelE}
\Delta E = \frac{(a+\delta +\sqrt{\delta(\delta+2a)}) \left(2M^2(X-1)-qQ\right)-aL}{2M(a+\delta +\sqrt{\delta(\delta+2a)})}
\ee 
To ensure that $\Delta E > 0$, one has to show that there exists a parameter $\delta$ which yields
\be\label{del.constraint}
2M^2 \left( {\sqrt {1 + \frac{{L + q\left( {Q + q/2} \right) - \delta M}}{{M^2 }}}  - 1} \right) - qQ - \frac{{aL}}{{a + \delta  + \sqrt {\delta \left( {\delta  + 2a} \right)} }} > 0\,.
\ee 
Finding an exact range for $\delta$ which obeys the last inequality is kind of cumbersome job to do. Nevertheless, we still can show that $\Delta E > 0$ for $\delta >0$ without really solving the inequality (\ref{del.constraint}). From a minoration of (\ref{del.constraint}) which reads
\be\label{del.constraint.2}
2M^2 \left( {\sqrt {1 + \frac{{L + q\left( {Q + q/2} \right) - \delta M}}{{M^2 }}}  - 1} \right) - qQ - L > 0\,,
\ee 
one can find an upper limit for $\delta$ which reads
\be\label{del.constraint.3}
\delta  < \frac{{2q^2 J - L\left( {L + 2qQ} \right)}}{{2M\left( {2M^2  - q^2 } \right)}}
\ee 
which should also hold in (\ref{del.constraint}). Hence, from the equations (\ref{eqDelE}) and (\ref{del.constraint.3}) we learn that it is possible for a near-extremal Kerr-Sen black hole to be destroyed by test particles.

Now let us study the result above for the case of extremal Kerr-Sen black holes. Setting $\delta = 0$ in (\ref{eqDelE}) gives the range of $\Delta E$ in the extreme conditions as
\be \label{DelE.ext}
\Delta E = M\left[\sqrt{1+\frac{L+q(Q+q/2)}{M^2}}-1-\frac{L+qQ}{2M^2}\right]\,,
\ee
which could be positive when\footnote{This condition applies to the extremal case only.}
\be 
q > \frac{{L\left( {\sqrt 2 M + Q} \right)}}{{2J}}\,.
\ee

\begin{figure}
\begin{center}
\includegraphics[scale=0.4]{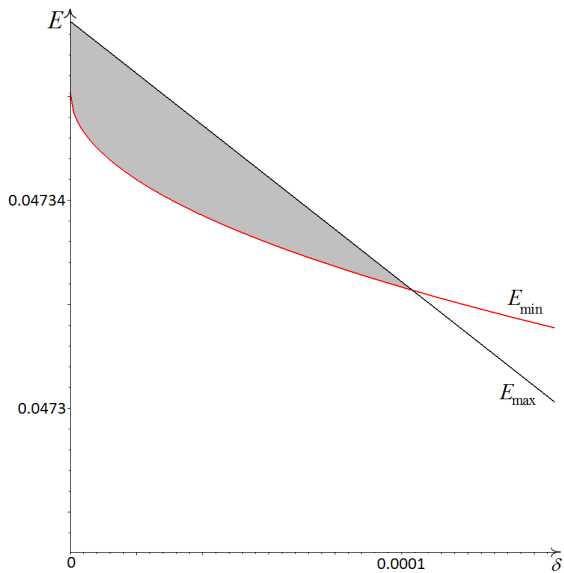}
\end{center}
\caption{Plot of $E$ - $\delta$. The shaded area is the intersection of $E<E_{\rm max}$ and $E>E_{\rm min}$.} \label{Fig.DelE}
\end{figure}

Therefore, we can learn from Eqs. (\ref{eqDelE}) and (\ref{DelE.ext}) that there is a narrow range of test particle's energy, $\Delta E$, which allows the appearance of naked singularities from both near-extremal and extremal Kerr-Sen black holes respectively. However, the particle's energy must be very finely tuned to yield the black hole's destruction, even if the particle is released from very far away. In supporting this finding, we provide Fig. \ref{Fig.DelE} which shows the plots of $E_{\rm max}$ and $E_{\rm min}$ where the choices of numerical values are $M=100$, $a=90$, $q=0.1$, and $L=5$. We find that the shaded area in Fig. \ref{Fig.DelE}, which represents the dependency of $\Delta E$ with respect to the parameter of near-extremality $\delta$, gets narrower as $\delta$ increases. Intuitively it means that the black hole becomes harder to break when it moves away from the extremality.

\begin{figure}
\begin{center}
\includegraphics[scale=0.4]{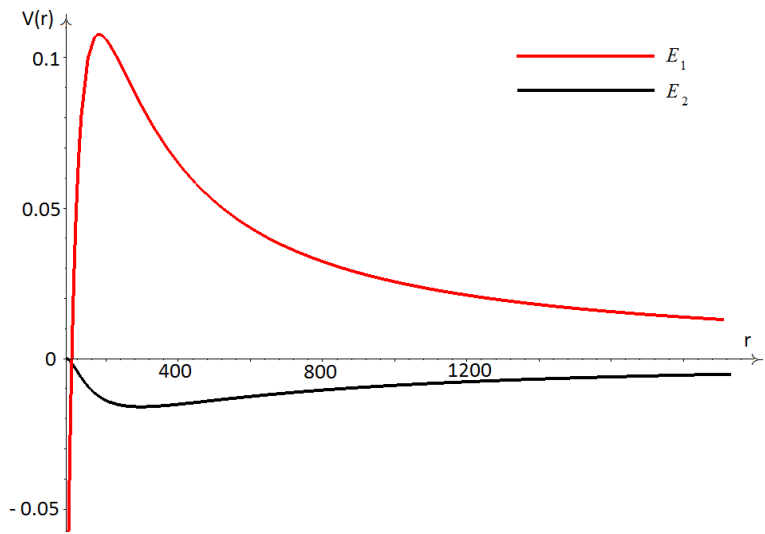}
\end{center}
\caption{The effective potential $V(r)$ of a test particle with several energies outside event horizon $r_h$, $E_1 < E_{\rm min}$ and $E_{\rm min}< E_2 < E_{\rm max}$.} \label{vr}
\end{figure}

It is interesting to note that the result\footnote{After performing the Taylor expansion to the root squared term in (\ref{DelE.ext}), one can show that $\Delta E \simeq q^2/4M$.} (\ref{DelE.ext}) is in accordance with one of the findings in \cite{Gao:2012ca}, where the authors pointed out that the range of particle's energy that could destroy an extremal Kerr-Newman black hole is of the order of $q^2/M$. Then, our next job is to verify that a particle from far away whose physical properties obey (\ref{eqDelE}) and (\ref{del.constraint}) could really fall into the black hole. A simple way to perform this task is by considering the particle's orbit which lies entirely on the $\theta = \pi /2$ plane\footnote{In the Appendix \ref{app.equator} we show explicitly that such a plane exists in the case of Kerr-Sen black hole.}. Related to the test particle described in Fig. \ref{Fig.DelE}, one can learn from Fig. \ref{vr} that a test particle whose energy is in between $E_{\rm max}$ and $E_{\rm min}$ could really jump into a Kerr-Sen black hole from infinity. 

The corresponding effective potential for the test particle plotted in Fig. \ref{vr} is obtained from Eq. (\ref{EtoL}) by using the relation $V(r) = -\dot r^2$ \cite{Gao:2012ca}. The numerical values which are used to produce this figure are the same as those in obtaining Fig. \ref{Fig.DelE}, i.e. $M=100$, $a=90$, $q=0.1$, and $L=5$, with $\delta$ is chosen to be $10^{-5}$. This choice of $\delta$ obeys the inequality (\ref{del.constraint.3}), which for the numerical values that lead to Fig. \ref{vr} reads $\delta < 2.757\times 10^{-5}$. Consequently, the resulting numerical values for $E_{\rm min}$ and $E_{\rm max}$ are $0.04734888626$ and $0.0473694$, respectively, and the energy $E$ of test particle which has the possibility to destroy the black hole must be within these energies, i.e. $E_{\rm min} < E < E_{\rm max}$. In Fig. \ref{vr}, the plot $E_1$ describes a particle with $E=m=0.04$, and the plot $E_2$ for a particle with $E=m=0.04736$.

\section{Static and neutral limits of Kerr-Sen black holes}\label{s.4}

It is known that the Kerr and GMGHS solutions can be obtained from Kerr-Sen spacetime by taking $Q\to 0$ and $a\to 0$ respectively. It resembles the case of Einstein-Maxwell theory, where one can get Kerr and RN appear when the limits $Q\to 0$ and $a\to 0$ are performed in the Kerr-Newman solution respectively. In this section, we present some analysis related to the possibility of transforming Kerr and GMGHS black holes into some naked singularities as an effect of test particle capture.

From the equations (\ref{E-max}) and (\ref{Emin-notEx}), the maximum and minimum particle energies which allow the possibility of producing naked singularities from Kerr black holes are
\be\label{EmaxKerr}
E_{\max }  = M\left( {\sqrt {1 + \frac{{L - \delta M}}{{M^2 }}}  - 1} \right)\,,
\ee 
and
\be \label{EminKerr}
E_{\min }  = \frac{{aL}}{{2M\left( {M + \sqrt {\delta \left( {M + a} \right)} } \right)}}\,.
\ee 
The last two formulas yield the range of energy for particle to have the possibilities to destroy a Kerr black hole:
\be\label{DEKerr}
\Delta E = \frac{{2M^2 \left( {M + \sqrt {\delta \left( {M + a} \right)} } \right)\left( {\sqrt {1 + \frac{{L - \delta M}}{{M^2 }}}  - 1} \right) - aL}}{{2M\left( {M + \sqrt {\delta \left( {M + a} \right)} } \right)}}\,.
\ee 
If one can show that there is a range of $\delta$ which yields the nonvanishing and positive $\Delta E$ in (\ref{DEKerr}), than we can conclude that the near-extremal Kerr black holes, or even the extremal ones depending on whether $\delta=0$ is covered by the range, can be destroyed by the test particle whose energy is within $E_{\rm max}$ and $E_{\rm min}$. However, performing such task is found to be not quite simple due to the nonlinearity of $\delta$ in the corresponding equation. Nevertheless, we can see that $E_{\rm min} > E_{\rm max}$ when the Kerr black holes are in extremality, which means an extremal Kerr black hole cannot be turned into a naked singularity by feeding it a test particle with angular momentum. On the other hand, in the range of $\delta$ that satisfies the inequality
\be
2M^2 \left( {M + \sqrt {\delta \left( {2M - \delta } \right)} } \right)\left( {\sqrt {1 + \frac{{L - \delta M}}{{M^2 }}}  - 1} \right) + \left( {\delta  - M} \right)L > 0\,
\ee 
one can find that $\Delta E >0$. Consequently, we can conclude that it is possible the near-extremal Kerr black holes to be destroyed by a test particle whose energy is very finely tuned within the range of $E_{\rm min} < E < E_{\rm max}$. 

\begin{figure}
\begin{center}
\includegraphics[scale=0.35]{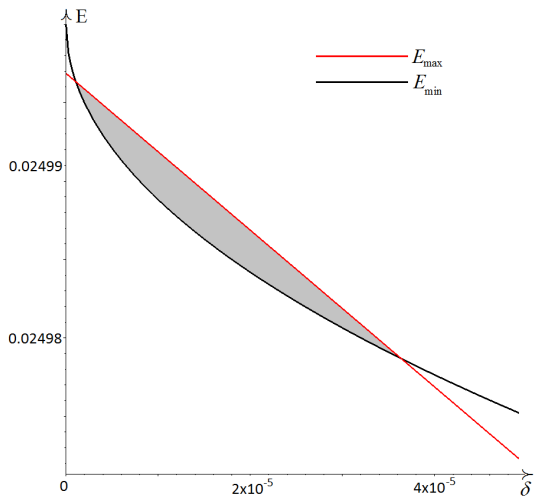}
\end{center}
\caption{The shaded area in the graphic above shows $\Delta E$ which allows naked singularity to be produced from the near-extremal Kerr black holes. When $\delta = 0$, $\Delta E < 0$, meaning that the minimum energy of the particle to arrive at the event horizon is higher than the maximum energy that may allow the violation of extremal bound for Kerr black holes.} \label{dEKerr.fig}
\end{figure}

To support the claim above, that a near-extremal Kerr black hole can be turned into a naked singularity, but not when it is already in the extremal condition, we provide a numerical analysis for $E_{\rm max}$ and $E_{\rm min}$ in Fig. \ref{dEKerr.fig}. The plots in Fig. \ref{dEKerr.fig} are obtained by setting $M=100$ and $L=5$, where we can see that there is a narrow range of $\delta$ which yields $\Delta E >0$. In this narrow range of $\delta$ black hole can be destroyed, but not when the black hole is extremal where it can be seen that $E_{\rm min} > E_{\rm max}$. This result is in agreement to that in \cite{Jacobson:2009kt} where the authors showed that the overspinning of Kerr black holes is possible if the black hole starts out nearly below the maximal spin. 

For the case of GMGHS black holes, one can set $a=0$ and $L=0$ in (\ref{E-max}) and (\ref{Emin-notEx}). This leads to
\be
\Delta E = E_{\max }  - E_{\min }  = M\left( {\sqrt {1 + \frac{{q\left( {Q + q/2} \right) - \delta M}}{{M^2 }}}  - \left( {1 + \frac{{qQ}}{{2M^2 }}} \right)} \right)
\ee 
which can never be nonzero positive for $q \ll Q$. Therefore, unlike the RN black hole which can be overcharged from below the extremality \cite{Hubeny:1998ga}, neither the near-extremal nor extremal GMGHS black holes can be overcharged to pass the extremal point. This is in full agreement with an argument by Horowitz based on the area theorem \cite{Horowitz92}. However, it is interesting to note that the event horizon's radius of GMGHS black holes shrinks to zero when the black holes reach the extremal condition. 

\section{Discussion and Conclusions}\label{s.discussion}

In this paper we have adopted the gedanken experiment by Wald to show the possibility of destroying a (near-)extremal Kerr-Sen black hole by feeding it a charged massive test particle. First we computed the minimum and maximum energies of the test particle, related to the situation where it has enough energy to get close to the event horizon, but not too energetic so the extremal condition for Kerr-Sen black holes can still be violated. We find that, by neglecting the radiative and self-force effects, a test particle can destroy a Kerr-Sen black hole, in both near-extremal and extremal conditions. When the limit $Q\to 0$ is taken in the expressions of $E_{\rm min}$ and $E_{\rm max}$ for the test particle, we find that only the near-extremal Kerr black holes can be destroyed, in agreement with the work by Jacobson and Sotiriou \cite{Jacobson:2009kt}. On the other hand, taking $a\to 0$ in the corresponding $E_{\rm min}$ and $E_{\rm max}$ shows that the GMGHS black holes can never be overcharged. This supports the conclusion by Horowitz for GMGHS black holes based on the area theorem \cite{Horowitz92}. 

Nevertheless, the analysis performed in this paper neglects the radiative and self-force effects. The fact that $\Delta E$ is very small leaves the possibility of radiative and self-force effects\ considerations to cure the problem of WCCC violation in Kerr-Sen spacetime. Also one might raise a question about the chance of turning a Kerr-Sen black hole into a naked singularity if the test particle has charge related to the antisymmetric tensor field $B_{\mu\nu}$. We will address these projects in our future work.

\section*{Acknowledgements}

I thank R. Primulando for reading this manuscript. I also thank K. Virbhadra, I. Semiz, and the anonymous referee for their useful comments and suggestions.

\appendix
\section{Equatorial plane of Kerr-Sen spacetime}\label{app.equator}

Related to the Lagrangian (\ref{eq.Lang}), one can show
\[
\frac{{\partial {\cal L}}}{{\partial \theta }} = \frac{{ma^2 \sin 2\theta }}{2}\left( {\frac{{2Mr}}{{\rho ^2 }}\dot t^2  - \frac{{\dot r^2 }}{\Delta } - \dot \theta ^2  + \frac{\Xi }{{a^2 \rho ^4 }}\dot \phi ^2  - \frac{{4Mr\left( {\Delta  + 2Mr} \right)}}{{a\rho ^4 }}\dot t\dot \phi } \right)
\]
\be 
 + qrQ\sin 2\theta \left( { - \frac{{a^2 }}{{\rho ^4 }}\dot t + \frac{{a\left( {\Delta  + 2Mr} \right)}}{{\rho ^4 }}\dot \phi } \right)\,,
\ee
where
\[\Xi = 2rM{a}^{2} \left( {r}^{2}+2rb+{a}^{2} \right)  \sin^2 \theta+\Delta{a}^{4} \cos^4 \theta+2r{a}^{2} \left( M{a}^{2}+{r}^{2}M+2\Delta
b+2rMb+r\Delta \right)  \cos^2 \theta \]
\[+{r}^{2} \left( r+2b \right)  \left( 2{r}^{2}M+r
\Delta+4rMb+2\Delta b+2M{a}^{2} \right) 
\]
and 
\be 
p_\theta   \equiv \frac{{\partial {\cal L}}}{{\partial \dot \theta }} = m\rho ^2 \dot \theta \,.
\ee
Hence it is obvious that the particle's motion on the equatorial plane, i.e. fixed $\theta = \pi/2$, satisfies the Euler-Lagrange equation
\be 
\frac{\partial {\cal L}}{\partial \theta} = {\dot p_\theta}\,.
\ee

\end{document}